\documentclass{ws-procs9x6}            

\newcommand{\compass}{\textsc{Compass}}

\begin{document}
\title{Freed-Isobar Analysis of Light Mesons at COMPASS}

\author{F. Krinner$^*$ for the \compass\ collaboration}

\address{Max Planck Institut for Physics,\\
80805 Munich, Bavaria, Germany\\
$^*$E-mail: fkrinner@mpp.mpg/de}

\begin{abstract}
Modern hadron-spectroscopy experiments such as \compass\ collect data samples 
of unprecedented size, so that novel analysis techniques become possible and 
necessary. One such technique is the freed-isobar partial-wave analysis (PWA). 
In this approach, fixed parametrizations for the amplitudes of intermediate 
states commonly modeled using Breit-Wigner shapes are replaced by sets of 
step-like functions that are determined from the data. This approach not only 
reduces the model dependence of partial-wave analyses, but also allows us to 
study the amplitudes of the intermediate states and their dependence on the 
parent system.

We will also present results of a freed-isobar PWA performed on the large data 
set on diffractive production of three charged pions collected by the \compass\ 
experiment, which consists of $46\times10^6$ exclusive events. We will focus on 
results for the wave with spin-exotic quantum numbers $J^{PC}=1^{-+}$, in 
particular on its decay into $\rho(770)\pi$. Here, the freed-isobar PWA 
method provides insight into the interplay of three- and two-particle dynamics
and confirms the decay of the spin-exotic $\pi_1(1600)$ resonance to 
$\rho(770)\pi$ in a model-independent way.
\end{abstract}


\bodymatter

\section{Diffractive $3\pi$ production}\label{aba:sec1}
The data set presented here was collected in 2008 by the \compass\ experiment 
located at CERN's North Area using a $190\,\text{GeV}/c$ $
\pi^-$ beam impinging on liquid hydrogen serving as a proton target. Using the 
two-stage \compass\ spectrometer, the process
\begin{equation}
 \pi^-_\text{beam} + p_\text{target} \to \pi^-\pi^+\pi^- + p_\text{recoil}
\end{equation}
is selected resulting in a sample of $46\times10^6$ exclusive events. These 
events have been analyzed in an extensive PWA \cite{massIndep} and the 
resonance parameters of eleven isovector $3\pi$ resonances have been 
extracted \cite{massDep}.

\section{Freed-isobar PWA method}
The measured intensity distribution $\mathcal{I}(\vec\tau)$ for a given $m_{3\pi}$ 
and $t^\prime$ bin is a function of the five phase 
space variables $\vec\tau$ and is modeled as the modulus square of a coherent sum 
over partial-wave amplitudes:
\begin{equation}
 \mathcal{I}(\vec\tau)=\left|\sum_\text{waves} \mathcal{T}_\text{wave}\mathcal{A}_\text{wave}(\vec\tau)\right|^2
.\end{equation}
Here, the complex-valued transition amplitudes $\mathcal{T}_\text{wave}$ encode 
the strengths and phases with which single partial waves contribute and the 
decay amplitudes $\mathcal{A}_\text{wave}(\vec\tau)$ describe the $3\pi$ decay and encode 
the $\vec\tau$-dependence of the partial waves.

The conventional PWA model uses a set of 88 partial waves that 
differ in spin quantum numbers and $2\pi$ resonance content\cite{massIndep}. The decay 
amplitudes are split up under the assumption that production 
and decay of appearing $2\pi$ and $3\pi$ resonances factorize and the process can therefore 
be described as two subsequent two-particle decays. This assumption is known as the 
isobar model:
\begin{equation}
 \mathcal{A}_\text{wave}(\vec\tau)=\psi_\text{wave}(\vec\tau)\Delta_\text{wave}(m_{\pi\pi}) + \text{Bose symm.}
,\end{equation}
where $\psi_\text{wave}(\vec\tau)$ describes the dependence of the partial wave 
on the decay angles which is fully determined from first principles by the 
appearing spin quantum numbers. The dynamic isobar amplitudes 
$\Delta_\text{wave}(m_{\pi\pi})$ in contrast describe the $2\pi$ resonance 
content---or isobar---of the partial waves and have to be known beforehand. 
Bose symmetrization is necessary due to the two identical final-state $\pi^-$.

The choice of $2\pi$ resonance content, its parameterizations and its parameters 
introduces possible model bias to the partial wave model and neglects possible 
contributions from low-intensity resonances or final-state interactions of the 
isobar with the third pion. To overcome this, we re-analyzed the data set using 
the freed-isobar approach, in which we replace the fixed dynamic isobar 
amplitudes by sets of indicator functions spanning the kinematically allowed 
$m_{\pi\pi}$ range:
\begin{equation}\label{eq::repl}
 \Delta_\text{wave}\big(m_{\pi\pi}\big)\to\sum_\text{bins}\Delta_\text{wave}^\text{bin}\big(m_{\pi\pi}\big)
 \quad\text{with}\quad \Delta^\text{bin}_\text{wave}\big(m_{\pi\pi}\big)=\begin{cases}1 & \text{$m_{\pi\pi}\in$ bin},\\0 & \text{otherwise.}\end{cases}
\end{equation}
This replacement alleviates the necessity for fixed dynamic isobar amplitudes from the partial
wave model, resulting in a much higher number of parameters, that can lead to
mathematical ambiguities within the model---so-called zero modes---that have to be 
identified and resolved. The detailed origin of these ambiguities in 
general and for the data presented here is discussed in  \begingroup
    \romannumeral-`\x 
    \setcitestyle{numbers}%
    Refs.~[\cite{zmPaper,meson,PHD}].%
  \endgroup

We replaced the dynamic isobar parameterizations of 12 of the 88 waves in the 
PWA model of \begingroup
    \romannumeral-`\x 
    \setcitestyle{numbers}%
    Ref.~[\cite{massIndep}] \endgroup following eq.~\ref{eq::repl}. Eleven of these waves
 represent the waves describing the largest intensities in the model, so 
that 75\% of the total intensity is described by the freed waves. This minimizes possible effects of 
imperfections in the fixed isobar parameterizations of the remaining waves. The twelfth wave is the 
spin-exotic $1^{-+}1^+[\pi\pi]_{1^{--}}\pi\text{P}$ wave\footnote{The wave name is given by $J^{PC}_{3\pi}M^\varepsilon[\pi\pi]_{J^{PC}_{\pi\pi}}\pi\text{L}$, 
          where $J^{PC}$ are the quantum number of the $3\pi$ and $\pi\pi$ 
          systems, $M^\varepsilon$ is the spin-projection and reflectivity 
          quantum number of the $3\pi$ system and $\text{L}$ is the orbital 
          angular between the $\pi\pi$ system and the third pion.}, which will be presented in Sec.~\ref{sec::result}. 
We use an $m_{\pi\pi}$ bin width of $40\,\text{MeV}/c^2$, and a region of narrower bins of 
$20\,\text{MeV}/c^2$ around the $\rho(770)$ resonance. 
The analysis was performed in 50 
equidistant bins in the mass range $0.5<m_{3\pi}<2.5\text{GeV}/c^2$ 
and four non-equidistant bins in the squared four-momentum transfer $t^\prime$, 
resulting in 200 independently fitted kinematic cells.
\begin{figure}[p]
\begin{center}
\includegraphics[width=2.1125in]{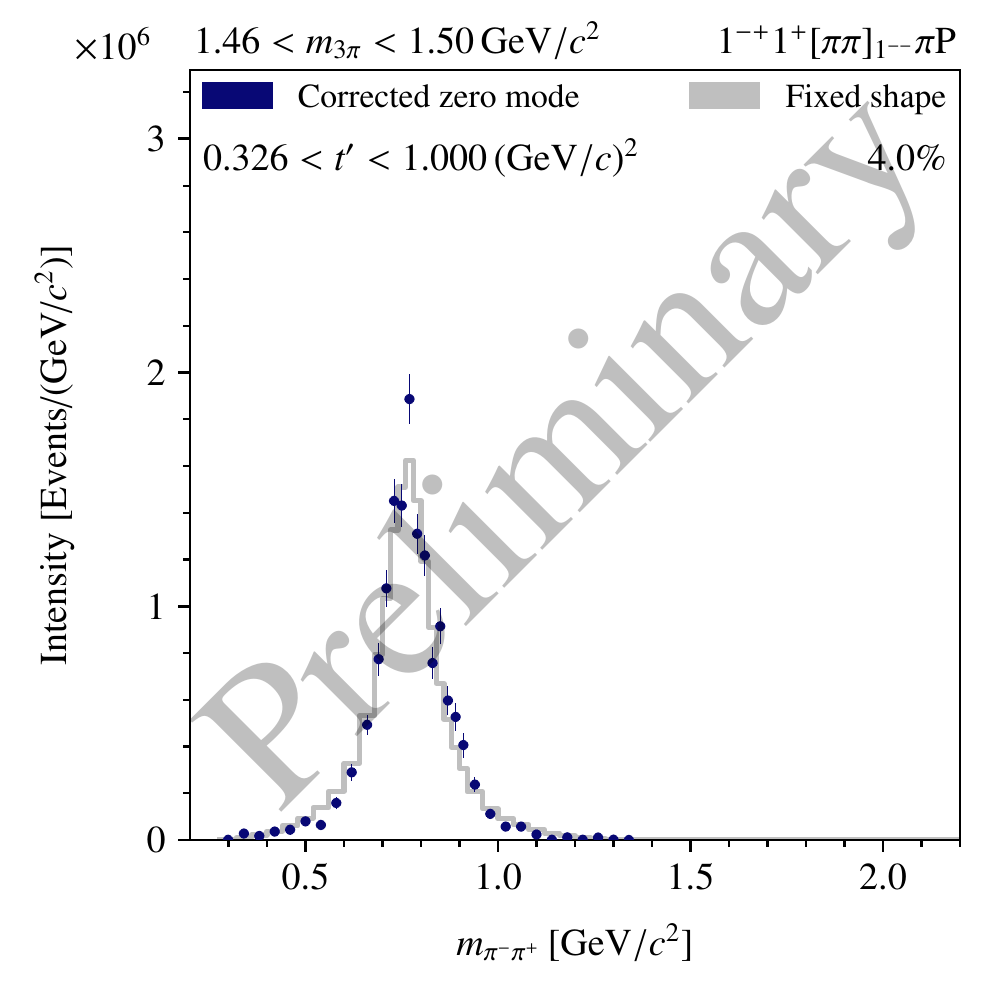}
\includegraphics[width=2.1125in]{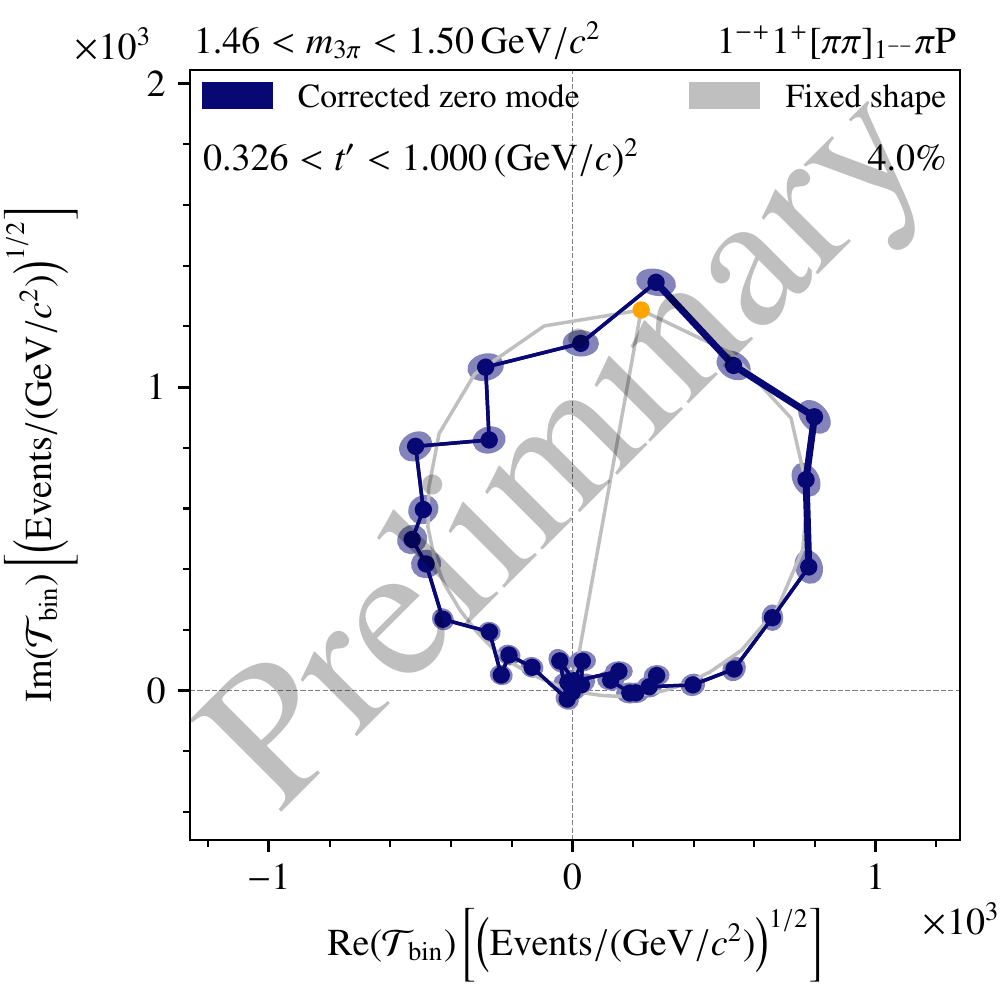}
\includegraphics[width=2.1125in]{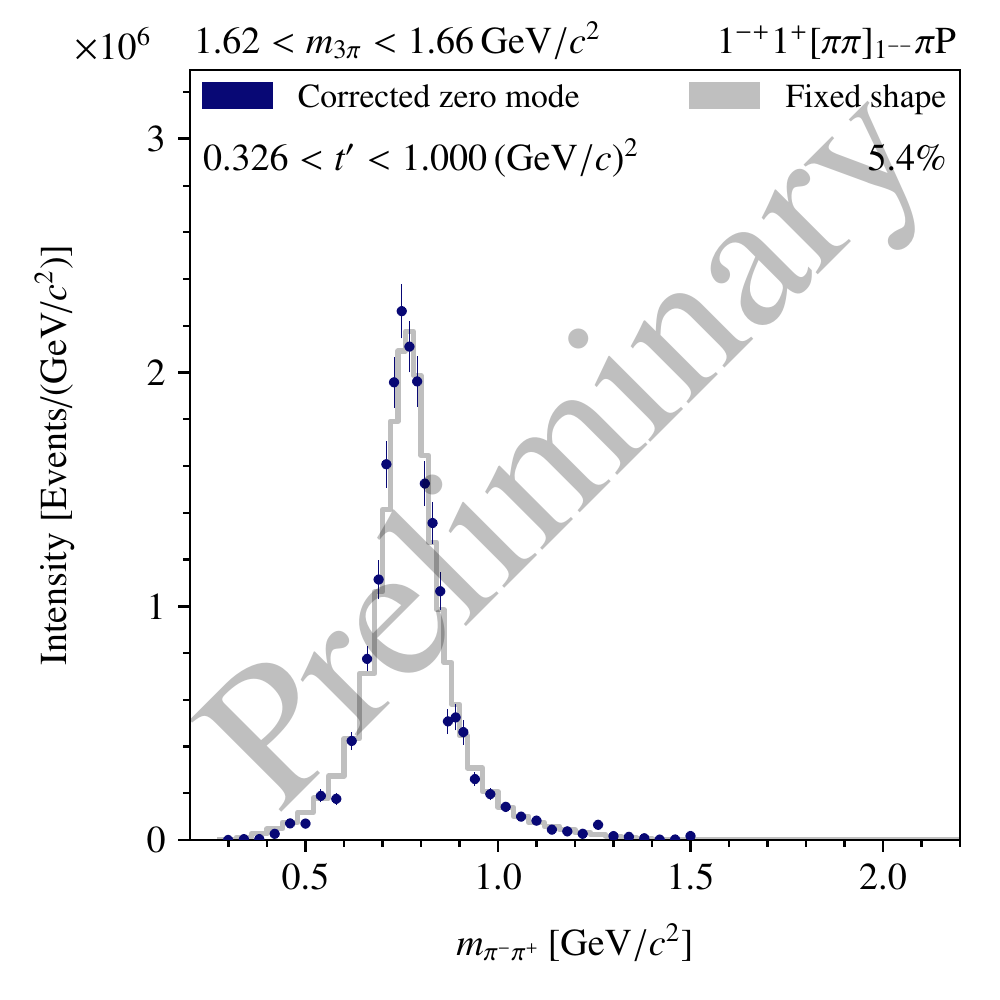}
\includegraphics[width=2.1125in]{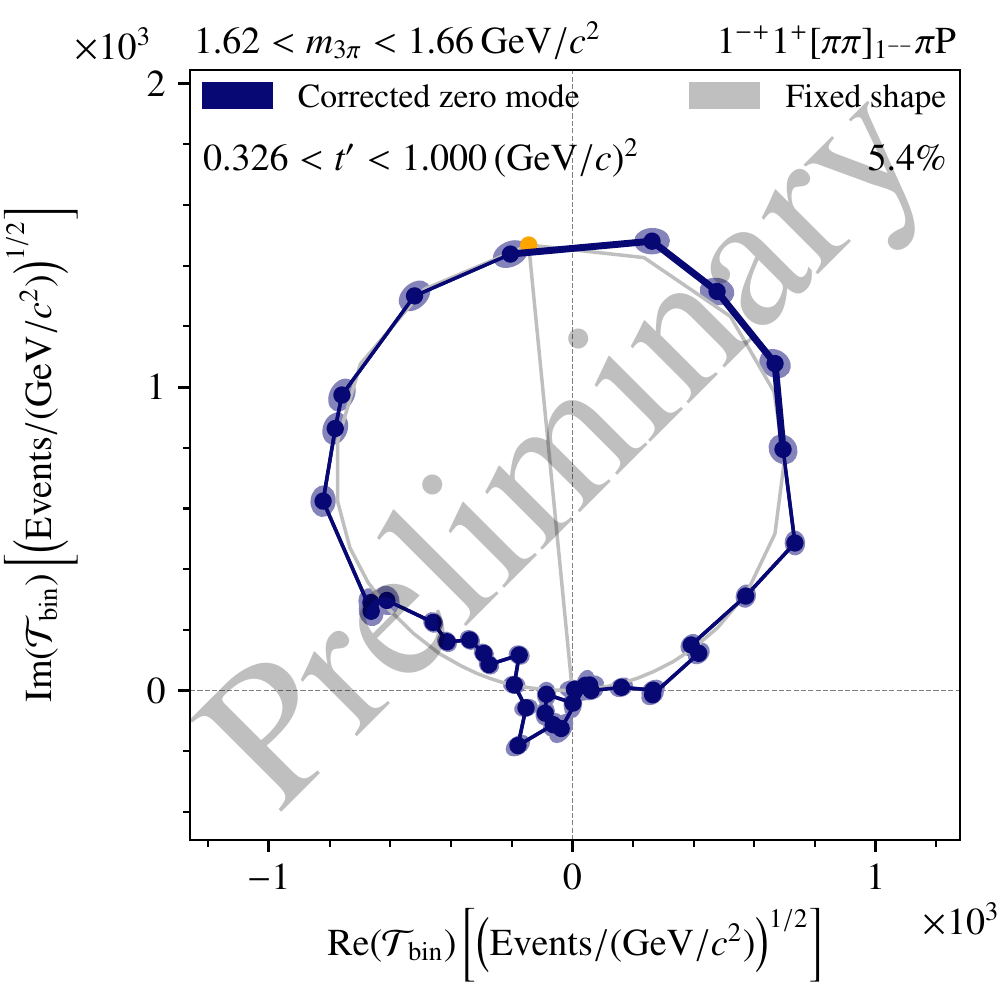}
\includegraphics[width=2.1125in]{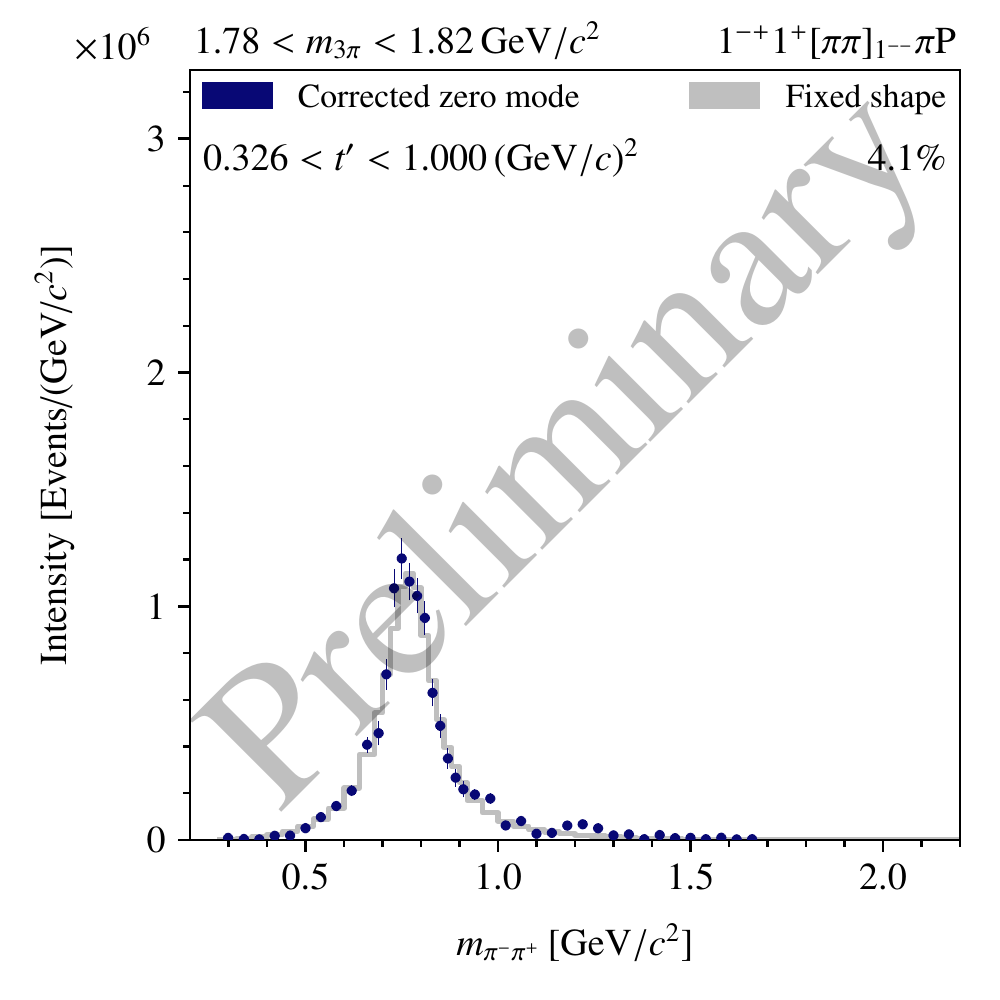}
\includegraphics[width=2.1125in]{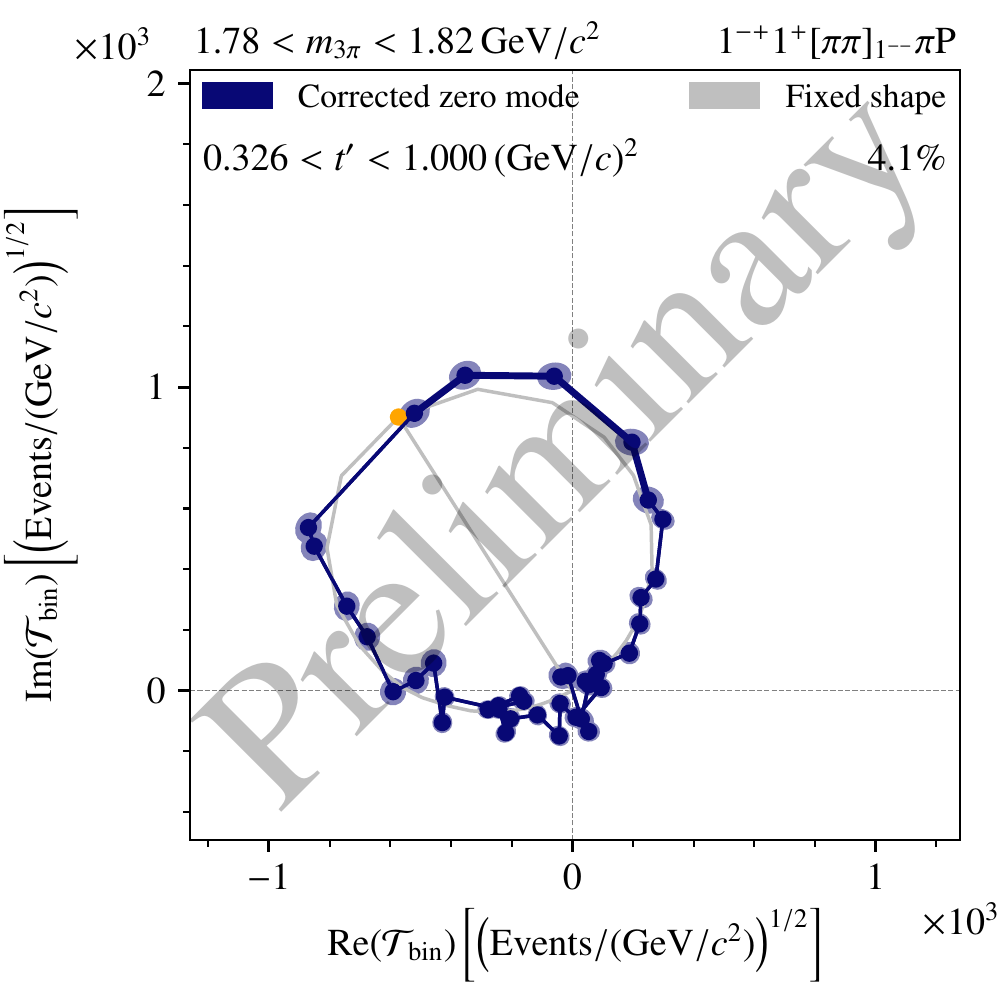}
\end{center}
\caption{Results of the freed-isobar PWA for three $m_{3\pi}$ bins (rows) in the highest $t^\prime$ bin  
  for the $1^{-+}1^+[\pi\pi]_{1^{--}}\pi\text{P}$ wave (blue). The Breit-Wigner amplitude for the $\rho(770)$
  as used in the conventional PWA is overlaid in gray. 
  Left: Intensity distributions. Right:~Argand diagrams; the orange dot indicates 
  the nominal resonance position of the gray curve.}
\label{fig::slices}
\end{figure}

\section{Results for the spin-exotic wave}
\label{sec::result}

\begin{figure}[t]
\begin{center}
\includegraphics[width=1.9in]{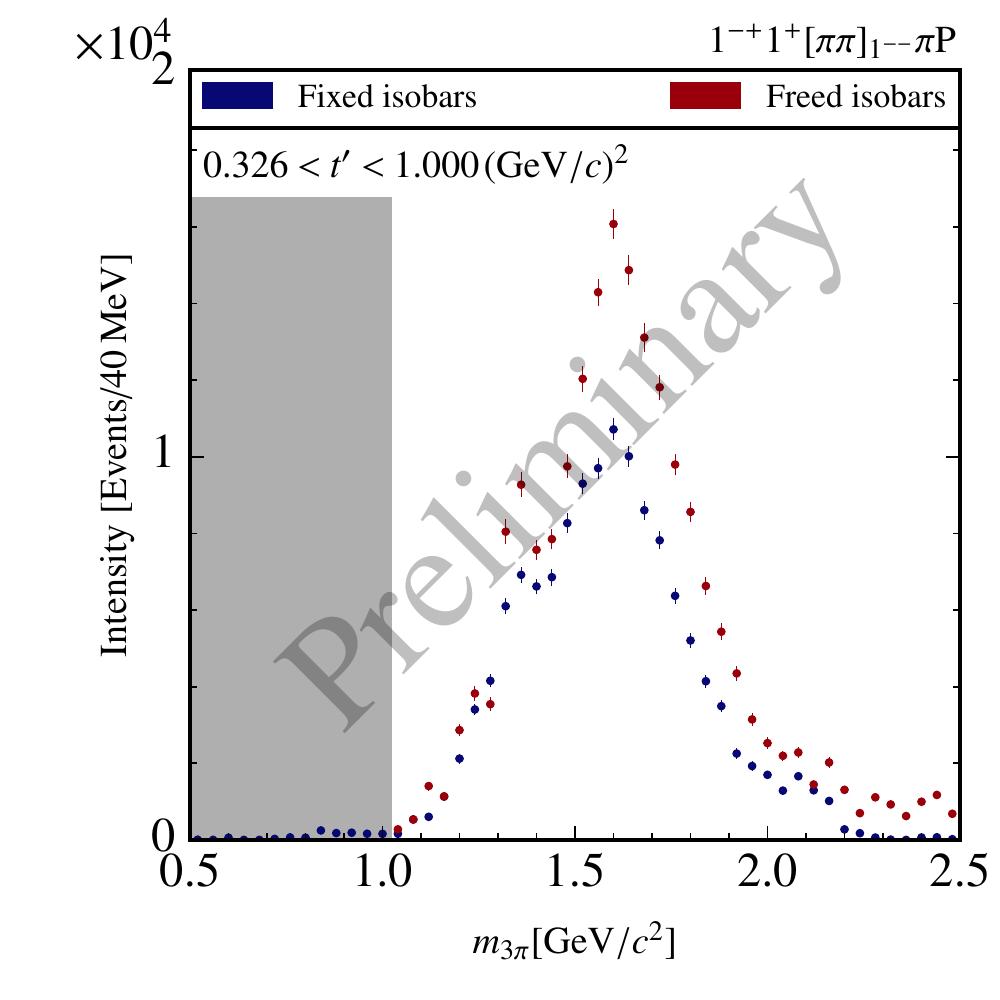}
\includegraphics[width=2.325in]{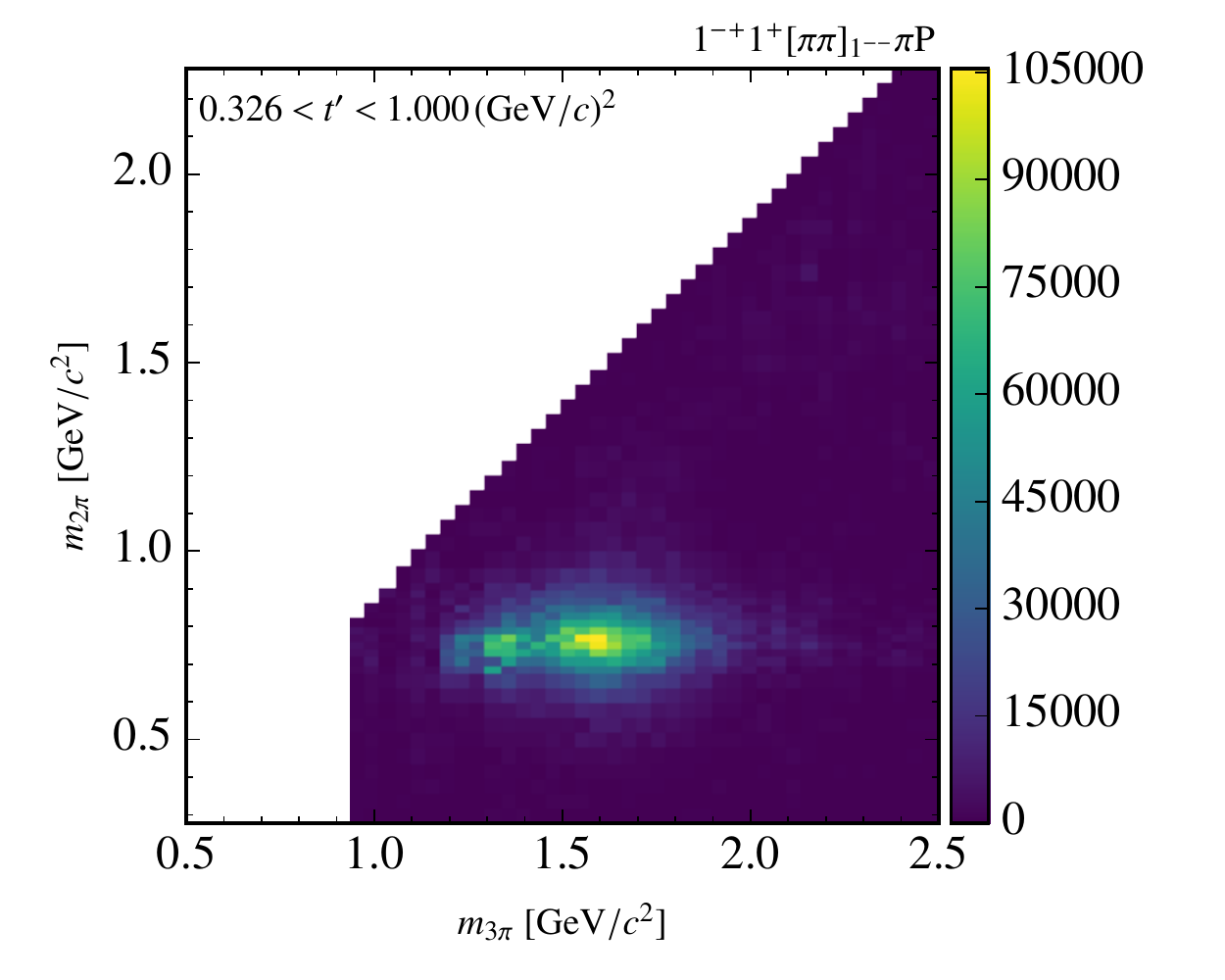}
\end{center}
\caption{Results of the freed-isobar analysis for the $1^{-+}1^+[\pi\pi]_{1^{--}}
  \pi\text{P}$ wave in the highest $t^\prime$ bin. Left: coherent sum over all 
  $m_{\pi\pi}$ bins (red) compared to the corresponding result of the conventional 
  PWA (blue). Right: Two-dimensional intensity distribution as function of $m_{3\pi}$ 
  and $m_{\pi\pi}$.}
\label{fig::3pispec}
\begin{center}
\includegraphics[width=2.1125in]{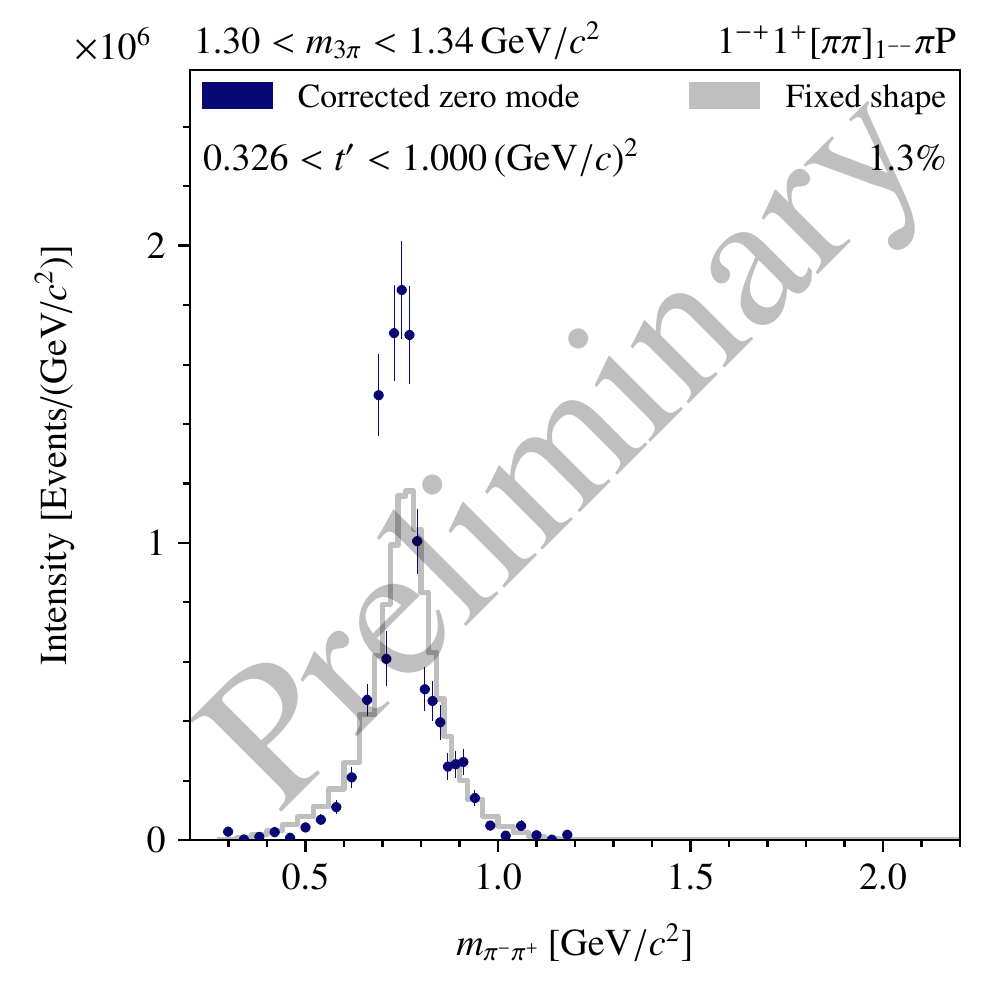}
\includegraphics[width=2.1125in]{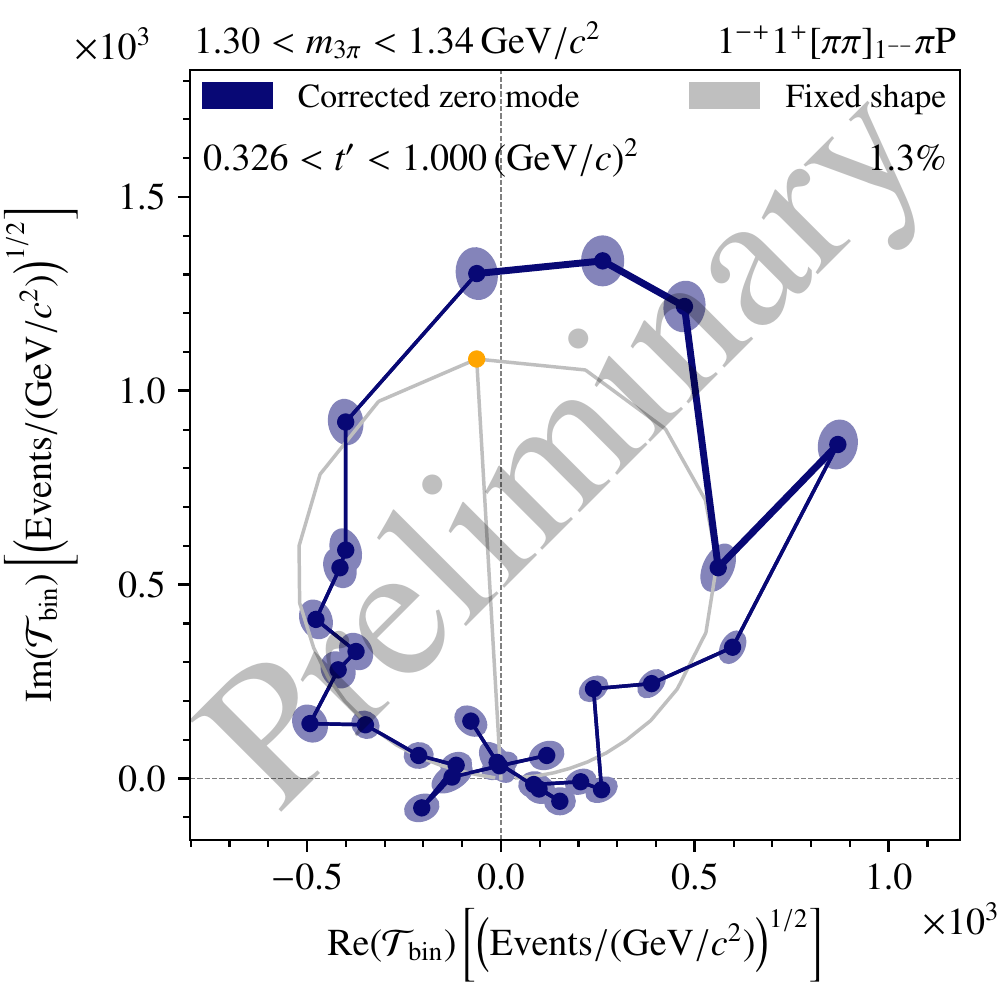}
\includegraphics[width=2.1125in]{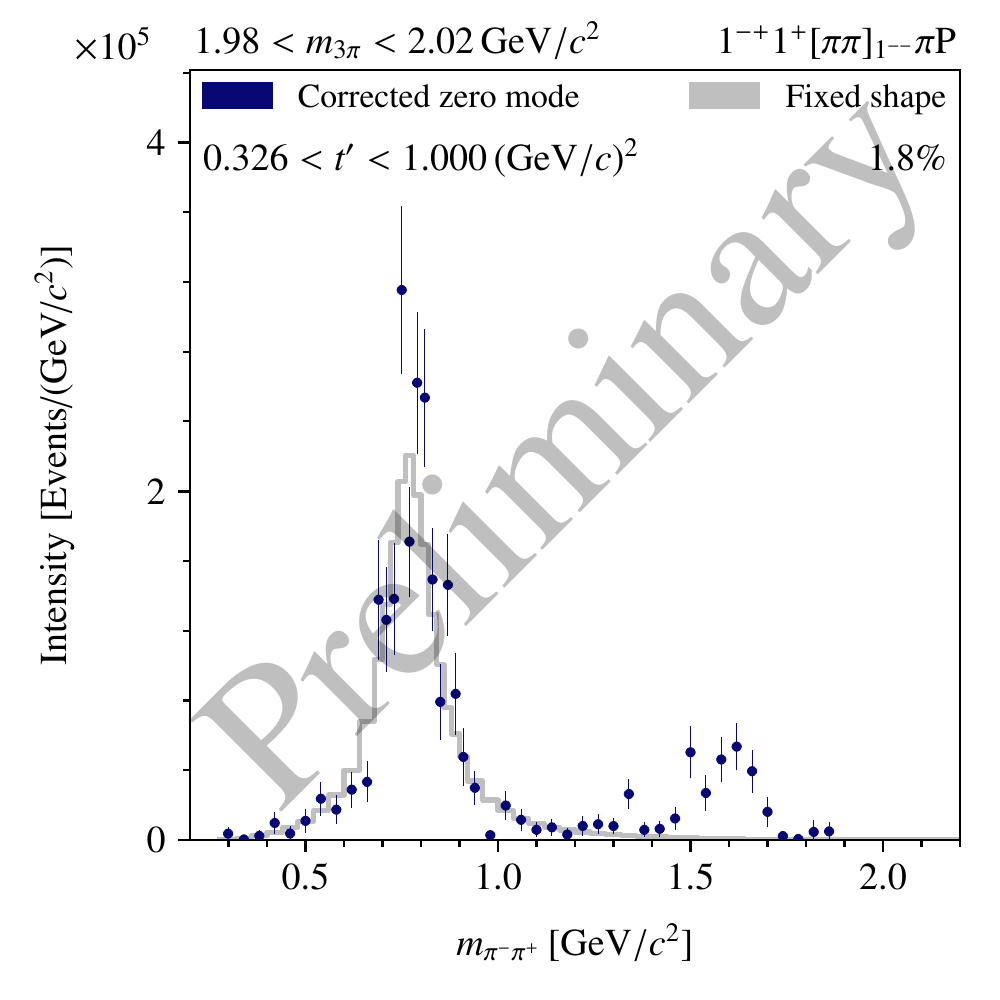}
\includegraphics[width=2.1125in]{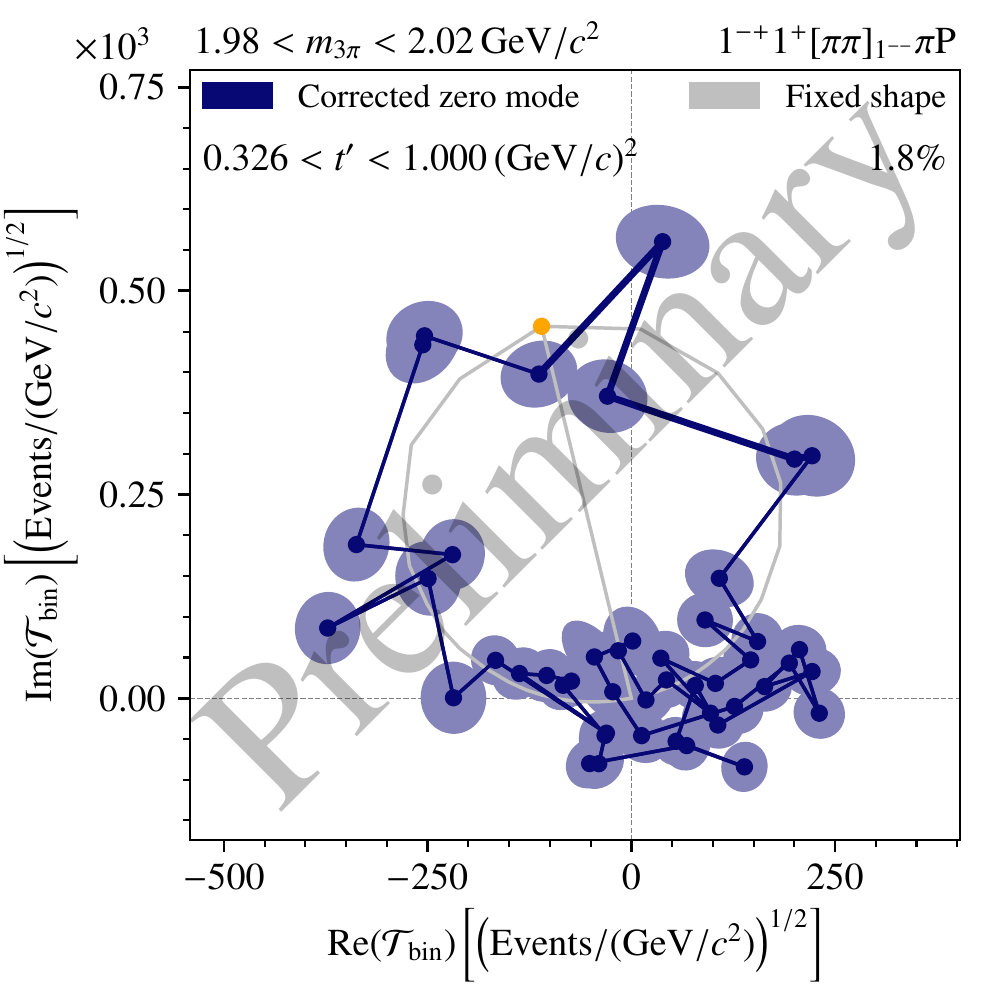}
\end{center}
\caption{As Fig.~\ref{fig::slices} for two additional $m_{3\pi}$ bins far away from the $\pi_1(1600)$ region.}
\label{fig::slicesToo}
\end{figure}

The result of this analysis is shown in Fig.~\ref{fig::3pispec}. As expected, the 
resulting 
two-dimensional intensity distribution on the right side is dominated by a clear
 peak corresponding to the decay $\pi_1(1600)\to\rho(770)\pi$. This
 confirms the existence of this decay without any assumption on the dynamic 
amplitude of the $[\pi\pi]_{1^{--}}$  system. The left plot shows the coherent 
sum of all $m_{\pi\pi}$ bins, compared to the result from the 
conventional PWA. Both distributions are dominated by a similar peak 
corresponding to the $\pi_1(1600)$ resonance.
The freed-isobar result has a higher over-all 
intensity, since it also can pick up intensity away from the $\rho(770)$ peak. 

Fig.~\ref{fig::slices} shows the results of the analysis of the data in the highest 
$t^\prime$ bin for three $m_{3\pi}$ bins, below, on, and above the nominal mass 
of the $\pi_1(1600)$ resonance. 
The left column shows the resulting 
intensity distributions and the right column shows the same data as Argand 
diagrams. In all three $m_{3\pi}$ bins, the fixed Breit-Wigner amplitude from the conventional 
PWA agrees nicely with the result of the freed-isobar analysis.

The intensity distribution assumes its highest value in the
$m_{3\pi}$ at the $\pi_1(1600)$ mass and the Argand diagrams rotate counter-clockwise moving through
the $\pi_1(1600)$ resonance, which both reflects the dynamic amplitude of the $\pi_1(1600)$ mother resonance, 
which further confirms its existence.

%

Since the fixed Breit-Wigner amplitude is adequate for all $m_{3\pi}$ slices in 
Fig.~\ref{fig::slices}, we show two additional bins in 
Fig.~\ref{fig::slicesToo} where the shapes do not match equally well. 
At low $m_{3\pi}$ the dynamic isobar amplitude, exhibits a 
sharper peak than the fixed Breit-Wigner, i.e. a slightly deformed Argand 
circle. At high $m_{3\pi}$, it exhibits a second peak in the 
intensity spectrum at a two-pion mass of $1.6\,\text{GeV}/c^2$, which could 
correspond to an excited $\rho^\prime$ and is unaccounted for in the 
conventional approach. Whether this peak actually stems from a resonance, 
requires a subsequent resonance-model fit.

\section{Conlusion}

We showed results for the dynamic isobar amplitudes of the spin-exotic 
$1^{-+}1^+[\pi\pi]_{1^{--}}\pi\text{P}$ wave extracted from \compass\ data on 
diffractively produced $3\pi$ using the freed-isobar approach \cite{zmPaper,meson,PHD}.
The extracted dynamic isobar amplitudes are in general in agreement with the 
assumption of a dominant presence of the $\rho(770)$ resonance in the $\pi\pi$ P-wave
wave as assumed in the conventional PWA approach\cite{massIndep}. In addition, 
the $\pi_1(1600)$ is confirmed to appear in this partial wave. Nevertheless, there are deviations from 
the Breit-Wigner amplitude that may originate from  distortions of the 
dynamic isobar amplitude due to non-resonant contributions, such as the Deck effect 
\cite{Deck} or re-scattering effects, or additional excited isobar resonances. To 
clarify the origin of these deviations, a resonance-model fit to the presented data is 
work in progress.

\small
\bibliographystyle{ws-procs9x6} 
\bibliography{proceedings_krinner}

\end{document}